\documentclass[doublecol]{myepl2}
\usepackage[centertags]{amsmath}
\usepackage{amssymb}
\usepackage{bm}
\usepackage{cite}
\newcommand{\vet}[1]{\ensuremath{\hskip-1pt\vec{\hskip1pt#1}}}

\title{Which is the Quantum Decay Law of Relativistic Particles?}
\shorttitle{Which is the Quantum Decay Law of Relativistic Particles?}

\author{S. A. Alavi\inst{1,2} \and C. Giunti\inst{2,3}}
\shortauthor{S. A. Alavi \and C. Giunti}

\institute{
\inst{1}
Department of Physics, Hakim Sabzevari University, P. O. Box, 397, Sabzevar, Iran
\\
\inst{2}
Department of Physics, University of Torino, Via P. Giuria 1, I--10125 Torino, Italy
\\
\inst{3}
INFN, Sezione di Torino, Via P. Giuria 1, I--10125 Torino, Italy
}


\abstract{
We discuss the relation between the
quantum-mechanical survival probability
of an unstable system in motion
and that of the system at rest.
The usual definition of the survival probability
which takes into account only the time evolution of an unstable system
leads to a relation
between the
survival probability
of the system in motion
and that of the system at rest
which is different from the standard relation
based on relativistic time dilation.
This approach led other authors to claim non-standard quantum-mechanical
effects which are in clear contradiction with Special Relativity.
We show that
an appropriate relativistic definition of the survival probability
which takes into account also the space evolution of an unstable system
leads to the standard relation
between the
survival probability
of the system in motion
and that of the system at rest,
in agreement with Special Relativity.
We present a rigorous derivation of this result based on a wave packet treatment.
}

\begin{document}

\maketitle

\section{Introduction}

It is well known that in Special Relativity
the decay lifetime of a system
in relativistic motion with velocity $\vet{v}$
is increased with respect to that of the same system at rest
by the factor\footnote{
We use natural units in which the velocity of light is equal to one.
}
$\gamma = \left( 1 - \vet{v}^2 \right)^{-1/2}$.
This law has been verified in countless experiments
(see, for example, Refs.~\cite{Frisch-Smith-1963,Bailey:1977de,Bailey:1978mn,Coan:2005st}).
However,
some authors
\cite{Shirokov:2004gi,Stefanovich:2005ai,quant-ph/0508087,physics/0603043,Urbanowski:2014gza}
obtained different results
from quantum mechanical calculations.
In this paper we show that a quantum mechanical treatment
which takes into account the space-time evolution of an unstable system
leads to the standard relativistic relation between
the decay lifetime of the system in relativistic motion
and that of the system at rest.

Let us emphasize that
the standard relativistic increase of the
decay lifetime of a system in motion by the factor $\gamma$
follows from elementary considerations
based on the relativistic time dilation
and on the physical requirement that
the fact that a system has decayed or not does not depend on its velocity with respect to the observer.
Indeed,
if we have an ensemble of systems at rest in a reference frame,
at every instant in time every observer counts the same
number of undecayed (or decayed) systems,
independently of the velocity of the observer with respect to the systems.
The authors of
Ref.~\cite{Shirokov:2004gi,Stefanovich:2005ai,quant-ph/0508087,physics/0603043,Urbanowski:2014gza}
seem to have a different opinion,
but for us this property of physical reality is obvious
and we can illustrate it with the following example.
Consider a muon passing through a detector,
where it can decay through the process
$\mu^{-} \to e^{-} + \bar\nu_{e} + \nu_{\mu}$.
It is well known that muons and electrons leave very different tracks,
which can be seen by all observers,
independently of their state of motion.
Hence, at every instant in time
all observers can see if the muon is decayed or not
and they all agree.

In the following we start with a brief review of the
quantum mechanical treatment of an unstable system in the rest frame,
then we discuss its extension to
a system in relativistic motion
and finally
we present a rigorous derivation of the survival probability
of an unstable system described by a wave packet.

\section{Unstable System at Rest}

In nonrelativistic quantum mechanics,
the probability that an unstable system at rest described by the state $|\Phi_{0}\rangle$ is not decayed at the time $t$ is
(see, for example, Ref.~\cite{Fonda:1978dk})
\begin{equation}
P_{0}(t)
=
|A_{0}(t)|^2
,
\label{e001}
\end{equation}
where $A_{0}(t)$ is the survival amplitude
\begin{equation}
A_{0}(t)
=
\langle \Phi_{0} | e^{-i\mathsf{H}t} | \Phi_{0} \rangle
,
\label{e002}
\end{equation}
where
$\mathsf{H}$
is the Hamiltonian operator.
An unstable system cannot have a definite energy,
because otherwise $P_{0}(t)=1$
and the system does not decay.
However, it is always possible to expand
the state $|\Phi_{0}\rangle$ over the eigenstates of the Hamiltonian.
Since the energy in the rest frame of the system is the mass of the system,
we denote the energy eigenstates by $|m\rangle$ and the energy eigenvalues by $m$,
such that
\begin{equation}
\mathsf{H} \, |m\rangle = m \, |m\rangle
.
\label{e003}
\end{equation}
The mass $m$ can assume a continuum of values
and the energy eigenstates are normalized by
\begin{equation}
\langle m | m' \rangle
=
\delta(m-m')
.
\label{e004}
\end{equation}
The expansion of the state $|\Phi_{0}\rangle$ over the eigenstates of the Hamiltonian is
\begin{equation}
| \Phi_{0} \rangle
=
\int dm \, \rho(m) \, |m\rangle
,
\label{e005}
\end{equation}
where
$\rho(m) = \langle m | \Phi_{0} \rangle$ and
\begin{equation}
\int dm \, |\rho(m)|^2 = 1
.
\label{e006}
\end{equation}
Here $|\rho(m)|^2$ is the distribution of mass
(energy in the rest frame) of the unstable system,
which determines the survival amplitude
\cite{Fock-Krylov-1947-JETP-17-93}.
From Eq.~(\ref{e002}) we obtain
\begin{equation}
A_{0}(t)
=
\int dm \, |\rho(m)|^2 \, e^{-imt}
.
\label{e007}
\end{equation}

Assuming a Breit-Wigner mass distribution
\begin{equation}
|\rho(m)|^2_{\text{BW}}
=
\frac{\Gamma/2\pi}{(m-M)^2+\Gamma^2/4}
,
\label{e008}
\end{equation}
where $M$ is the kinematical mass of the system
and
$\Gamma$
is the decay width,
and performing the integral in Eq.~(\ref{e007})
from $-\infty$ to $+\infty$,
one obtains the classical exponential decay law
\begin{equation}
P_{0}^{\text{BW}}(t)
=
e^{- \Gamma t}
.
\label{e009}
\end{equation}
However,
it is well known that the exponential decay law
is violated by all quantum systems
at small and large times.
In fact,
since any physical Hamiltonian must have a ground state which limits the amount of energy that can be extracted from the system,
the integral in Eq.~(\ref{e007})
has a lower bound which implies that at large times the
survival probability is a power law
\cite{Khalfin-1958-JETP-6-1053}.
Moreover,
if the mass distribution has a finite mean value\footnote{
Contrary to the Breit-Wigner mass distribution in Eq.~(\ref{e008})
whose mean value is not even defined.
},
the derivative of the survival probability at $t=0$
vanishes
(see Ref.~\cite{Fonda:1978dk}).

\section{Unstable System with Relativistic Velocity}

Let us now consider the same system in motion
with a relativistic velocity $\vet{v}$.
Elementary considerations based on relativistic time dilation lead to the standard relation
\begin{equation}
P_{\vet{v}}(t) = P_{0}(t/\gamma)
,
\label{e010}
\end{equation}
with $\gamma = \left( 1 - \vet{v}^2 \right)^{-1/2}$.
However,
some authors
\cite{Shirokov:2004gi,Stefanovich:2005ai,quant-ph/0508087,physics/0603043,Urbanowski:2014gza}
have used expressions of type (\ref{e002})
to calculate the survival probability of a relativistic system
and they have found relations which are different from Eq.~(\ref{e010})
and are in contradiction with Special Relativity.

In order to illustrate the problem,
let us consider the system in motion
with a relativistic velocity $\vet{v}$ and describe the
survival amplitude by
\begin{equation}
A_{\vet{v}}(t)
=
\langle \Phi_{\vet{v}} | e^{-i\mathsf{H}t} | \Phi_{\vet{v}} \rangle
.
\label{e011}
\end{equation}
Now we must take into account the momentum contribution to the
eigenstates of the Hamiltonian,
defined by
\begin{equation}
\mathsf{H} \, |E_{m}(\vet{p}),\vet{p},m\rangle = E_{m}(\vet{p}) \, |E_{m}(\vet{p}),\vet{p},m\rangle
,
\label{e012}
\end{equation}
where
\begin{equation}
E_{m}(\vet{p}) = \sqrt{\vet{p}^2 + m^2}
,
\label{e013}
\end{equation}
with the normalization
\begin{equation}
\langle E_{m}(\vet{p}),\vet{p},m | E_{m}(\vet{p}),\vet{p},m' \rangle
=
\delta(m-m')
.
\label{e014}
\end{equation}
In the rest frame, the unstable system is described by
\begin{equation}
| \Phi_{0} \rangle
=
\int dm \, \rho(m) \, |E_{m}(\vet{p})=m,\vet{p}=0,m\rangle
.
\label{e015}
\end{equation}
In a reference frame in which the system is in motion with velocity $\vet{v}$
we have
\begin{equation}
E_{m}(\vet{p}) = \gamma m
,
\qquad
\vet{p} = \gamma m \vet{v}
.
\label{e016}
\end{equation}
The states are transformed by the unitary operator
$\mathsf{U}_{\vet{v}}$,
such that
\begin{equation}
\mathsf{U}_{\vet{v}}
| E_{m}(\vet{p})=m, \vet{p}=0, m \rangle
=
| E_{m}(\vet{p})=\gamma m, \vet{p}=\gamma m \vet{v}, m \rangle
.
\label{e017}
\end{equation}
Therefore,
in this reference frame the system is described by
\begin{equation}
| \Phi_{\vet{v}} \rangle
=
\mathsf{U}_{\vet{v}}
| \Phi_{0} \rangle
=
\int dm \, \rho(m) \, |E_{m}(\vet{p})=\gamma m,\vet{p}=\gamma m \vet{v},m\rangle
.
\label{e018}
\end{equation}
Using this state,
from Eq.~(\ref{e011}) we obtain the survival amplitude
\begin{equation}
A_{\vet{v}}(t)
=
\int dm \, |\rho(m)|^2 \, e^{-im \gamma t}
,
\label{e019}
\end{equation}
which implies
\cite{quant-ph/0508087}
\begin{equation}
P_{\vet{v}}(t) = P_{0}(\gamma t)
.
\label{e020}
\end{equation}
This relation is completely different from the standard relation (\ref{e010}).

The reason why Eq.(\ref{e011}) gives a wrong result is
that it cannot describe the decay of a system in motion
with a relativistic velocity,
because the evolution operator
$e^{-i\mathsf{H}t}$
describes only the time evolution of the system,
whereas in Special Relativity
one must take into account both the time and space evolutions of a system,
which are perceived in different ways by different inertial observers.
These considerations lead us to consider the heuristic space-time dependent amplitude\footnote{
A rigorous treatment of the problem would require the use of
quantum field theoretical methods
(see Refs.~\cite{Nakazato:1995cn,Facchi:1999ik,Giacosa:2010br,Giacosa:2011xa}),
which are beyond the scope of the present work.
}
\begin{equation}
A_{\vet{v}}(t,\vet{x})
=
\langle \Phi_{\vet{v}} | e^{-i\mathsf{H}t+i\vet{\mathsf{P}}\cdot\vet{x}} | \Phi_{\vet{v}} \rangle
,
\label{e021}
\end{equation}
where $\vet{\mathsf{P}}$ is the momentum operator,
such that
\begin{equation}
\vet{\mathsf{P}} \, |E_{m}(\vet{p}),\vet{p},m\rangle = \vet{p} \, |E_{m}(\vet{p}),\vet{p},m\rangle
.
\label{e022}
\end{equation}
Apparently there is the problem that the amplitude (\ref{e021}) depends on the coordinate $\vet{x}$
whereas we are interested only in the survival probability as a function of the time $t$,
but we will see that we can overcome this problem.
Note however that there is no problem in the rest frame,
where $\vet{p}=0$
and
$A_{0}(t,\vet{x}) = A_{0}(t)$,
given by Eq.~(\ref{e002}).
In general,
using the expression (\ref{e018}) for
$| \Phi_{\vet{v}} \rangle$,
we obtain
\begin{equation}
A_{\vet{v}}(t,\vet{x})
=
\int dm \, |\rho(m)|^2 \, e^{-im \gamma (t - \vet{v}\cdot\vet{x})}
.
\label{e023}
\end{equation}
Now we notice that
since the system is moving with velocity $\vet{v}$,
its coordinate is given by
$\vet{x} = \vet{v} t$,
and we finally obtain the desired survival amplitude as a function of time only:
\begin{equation}
A_{\vet{v}}(t)
=
\int dm \, |\rho(m)|^2 \, e^{-imt/\gamma}
.
\label{e024}
\end{equation}
Confronting with the survival amplitude (\ref{e007}) of the system at rest,
one can see that
\begin{equation}
A_{\vet{v}}(t)
=
A_{0}(t/\gamma)
,
\label{e025}
\end{equation}
and
the standard relativistic relation (\ref{e010}) is satisfied.

The replacement
$\vet{x} = \vet{v} t$
is intuitively correct,
but a rigorous derivation
would be preferable.
Such a derivation can be obtained with the more complicated calculation
presented in the next Section,
in which the system is described by a wave packet.

\section{Wave Packet Treatment}

Let us first consider the wave packet describing the unstable system in the rest frame:
\begin{equation}
| \Phi_{0} \rangle
=
\int dm \, \rho(m)
\int d^3\vet{p}
\,
\varphi(\vet{p}) \, |E_{m}(\vet{p}),\vet{p},m\rangle
.
\label{e026}
\end{equation}
Since we must consider a continuum of values of the momentum $\vet{p}$,
the energy and momentum eigenstates are normalized by
\begin{equation}
\langle E,\vet{p},m | E',\vet{p}',m' \rangle
=
\delta^3(\vet{p}-\vet{p}')
\,
\delta(m-m')
,
\label{e027}
\end{equation}
and the momentum distribution
$\varphi(\vet{p})$
is normalized by
\begin{equation}
\int d^3\vet{p}
\,
|\varphi(\vet{p})|^2
=
1
,
\label{e028}
\end{equation}
in order to have $\langle \Phi_{0} | \Phi_{0} \rangle = 1$.

We consider a momentum distribution in the rest frame
such that\footnote{
For example the Gaussian momentum distribution
$
\varphi(\vet{p})
=
(2\pi)^{-3/4}
\sigma_{p}^{-3/2}
\exp\left( - \vet{p}^2 / 4 \sigma_{p}^2 \right)
$.
}
\begin{align}
\langle \vet{p} \rangle
=
\null & \null
\int d^3\vet{p}
\,
|\varphi(\vet{p})|^2
\vet{p}
=
0
,
\label{e029}
\\
\langle (p^{i})^2 \rangle
=
\null & \null
\int d^3\vet{p}
\,
|\varphi(\vet{p})|^2
(p^{i})^2
=
\sigma_{p}^2
\qquad
(i=1,2,3)
.
\label{e030}
\end{align}
The isotropic momentum uncertainty is assumed for simplicity,
whereas the vanishing average value
$\langle \vet{p} \rangle$
is required for a system at rest.
Moreover,
the system can be considered at rest only if
$\sigma_{p} \ll m$
for all the values of $m$ for which the mass distribution
$|\rho(m)|^2$
is not negligible,
i.e.
for
$M - \Gamma \lesssim m \lesssim M + \Gamma$,
where
$M$ is the average mass
and
$\Gamma$ is the decay width
(see for example the Breit-Wigner distribution in Eq.~(\ref{e008})).
Since the lifetime of an unstable system is well-defined and measurable
only if the decay width $\Gamma$ is much smaller than the mass $M$,
we have the condition
\begin{equation}
\sigma_{p} \ll M
.
\label{e031}
\end{equation}
This is an important condition which allows us to approximate
\begin{equation}
E_{m}(\vet{p})
\simeq
m
,
\label{e032}
\end{equation}
neglecting terms of order
$\vet{p}^2/m^2 \sim \sigma_{p}^2 / M^2 \ll 1$.

Let us now consider a reference frame in which the system is in motion with velocity $\vet{v}$
and let us call $\vet{k}_{m}$ the three-momentum in this frame,
in order to distinguish it from the three-momentum $\vet{p}$
in the rest frame.
The index $m$ in $\vet{k}_{m}$ is useful because
the value of $\vet{k}_{m}$ given by the Lorentz transformations
\begin{align}
E_{m}(\vet{k}_{m})
\null & \null
=
\gamma \left( E_{m}(\vet{p}) + v p_{\parallel} \right)
\simeq
\gamma \left( m + v p_{\parallel} \right)
,
\label{e033}
\\
k_{m\parallel}
\null & \null
=
\gamma \left( p_{\parallel} + v E_{m}(\vet{p}) \right)
\simeq
\gamma \left( p_{\parallel} + v m \right)
,
\label{e034}
\\
\vet{k}_{m\perp}
\null & \null
=
\vet{p}_{\perp}
,
\label{e035}
\end{align}
depends on $m$.
In Eqs.~(\ref{e033})--(\ref{e035})
$k_{m\parallel}$
($\vet{k}_{m\perp}$)
and
$p_{\parallel}$
($\vet{p}_{\perp}$)
are the components of
$\vet{k}_{m}$ and $\vet{p}$
parallel (orthogonal) to $\vet{v}$.
We have also adopted the approximation in Eq.~(\ref{e032}).

The energy and momentum eigenstates are transformed by
\begin{equation}
\mathsf{U}_{\vet{v}}
|E_{m}(\vet{p}),\vet{p},m\rangle
=
|E_{m}(\vet{k}_{m}),\vet{k}_{m},m\rangle
.
\label{e036}
\end{equation}
Therefore,
in the new reference frame the unstable system is described by
\begin{align}
| \Phi_{\vet{v}} \rangle
\null & \null
=
\mathsf{U}_{\vet{v}}
| \Phi_{0} \rangle
\nonumber
\\
\null & \null
=
\int dm \, \rho(m)
\int d^3\vet{p}
\,
\varphi(\vet{p}) \, |E_{m}(\vet{k}_{m}),\vet{k}_{m},m\rangle
.
\label{e037}
\end{align}
Note that since the operator $\mathsf{U}_{\vet{v}}$ is unitary,
we have
\begin{align}
\langle
E_{m}(\vet{k}_{m}),\vet{k}_{m},m
|
\null & \null
E_{m'}(\vet{k}'_{m'}),\vet{k'}_{m'},m'
\rangle
=
\langle E,\vet{p},m | E',\vet{p}',m' \rangle
\nonumber
\\
\null & \null
=
\delta^3(\vet{p}-\vet{p}')
\,
\delta(m-m')
,
\label{e0371}
\end{align}
and
$\langle \Phi_{\vet{v}} | \Phi_{\vet{v}} \rangle = 1$.

Using the state $| \Phi_{\vet{v}} \rangle$ in Eq.~(\ref{e037}),
from Eq.~(\ref{e021}) we obtain
the survival amplitude
\begin{equation}
A_{\vet{v}}(t,\vet{x})
=
\int dm \, |\rho(m)|^2
\int d^3\vet{p} \, |\varphi(\vet{p})|^2
\,
e^{-iE_{m}(\vet{k}_{m})t+i\vet{k}_{m}\cdot\vet{x}}
,
\label{e038}
\end{equation}
where
$E_{m}(\vet{k}_{m})$
and
$\vet{k}_{m}$
are the functions of $\vet{p}$ given by Eqs.~(\ref{e033})--(\ref{e035}).

The survival probability as a function of time is given by the normalized integral
over space of $|A_{\vet{v}}(t,\vet{x})|^2$:
\begin{equation}
P_{\vet{v}}(t)
=
\dfrac{
\int d^3x \, |A_{\vet{v}}(t,\vet{x})|^2
}{
\int d^3x \, |A_{\vet{v}}(t=0,\vet{x})|^2
}
.
\label{e039}
\end{equation}
The denominator is necessary for dimensional reasons and for the implementation
of the initial condition $P_{\vet{v}}(t=0)=1$.

The numerator in Eq.~(\ref{e039}) is given by
\begin{align}
\int d^3x
\null & \null
\,
|A_{\vet{v}}(t,\vet{x})|^2
=
\int dm \, |\rho(m)|^2
\int dm' \, |\rho(m')|^2
\nonumber
\\
\null & \null
\times
\int d^3\vet{p} |\varphi(\vet{p})|^2
\int d^3\vet{p}' |\varphi(\vet{p}')|^2
e^{-i[E_{m}(\vet{k}_{m})-E_{m'}(\vet{k}'_{m'})]t}
\nonumber
\\
\null & \null
\times
\int d^3x
\,
e^{i(\vet{k}_{m}-\vet{k}'_{m'})\cdot\vet{x}}
.
\label{e040}
\end{align}
The integral over $d^3x$ gives
\begin{align}
(2\pi)^3 \, \delta^3(
\vet{k}_{m}-\vet{k}'_{m'})
=
\null & \null
\frac{(2\pi)^3}{\gamma}
\,
\delta^2(\vet{p}_{\perp}-\vet{p}'_{\perp})
\nonumber
\\
\null & \null
\times
\delta(p_{\parallel} - p'_{\parallel} + v (m-m'))
,
\label{e41}
\end{align}
where we took into account Eq.~(\ref{e034}).
Using also Eq.~(\ref{e033}),
we obtain
\begin{align}
\int d^3x
\null & \null
\, |A_{\vet{v}}(t,\vet{x})|^2
=
\frac{(2\pi)^3}{\gamma}
\int dm \, |\rho(m)|^2
\nonumber
\\
\null & \null
\times
\int dm' \, |\rho(m')|^2
e^{-i(m-m')t/\gamma}
\nonumber
\\
\null & \null
\times
\int d^3\vet{p}
\left.
\,
|\varphi(\vet{p})|^2
\,
|\varphi(\vet{p}')|^2
\right|_{
\renewcommand{\arraystretch}{0.8}
\begin{array}{l} \scriptstyle
\vet{p}'_{\perp}=\vet{p}_{\perp}
\\ \scriptstyle
p'_{\parallel}=p_{\parallel}+v(m-m')
\end{array}
}
.
\label{e42}
\end{align}
The dependence on $v$ of the last term in Eq.~(\ref{e42})
spoils the derivation of a survival probability which satisfies the relativistic relation (\ref{e010}).
However,
we note that the lifetime $\tau$ of an unstable system is measurable only if the system
is localized in a region of space with an uncertainty $\sigma_{x} \ll \tau$.
In fact,
the fastest signal reaching the observer is a light signal,
whose time of emission has an uncertainty $\sigma_{x}$.
Since
$\sigma_{x}\sim\sigma_{p}^{-1}$
and
$\tau=\Gamma^{-1}$,
where $\Gamma$ is the width of the mass distribution
(see, for example, the Breit-Wigner distribution in Eq.~(\ref{e008})),
we have the condition
\begin{equation}
\Gamma \ll \sigma_{p}
.
\label{e421}
\end{equation}
In this case,
the difference $m-m'$, which is of order $\Gamma$,
is much smaller than the size $\sigma_{p}$ of the wave packet
and we can approximate Eq.~(\ref{e42}) with
\begin{align}
\int d^3x
\null & \null
\,
|A_{\vet{v}}(t,\vet{x})|^2
=
\frac{(2\pi)^3}{\gamma}
\left(
\int d^3\vet{p}
\,
|\varphi(\vet{p})|^4
\right)
\nonumber
\\
\null & \null
\times
\int dm \, |\rho(m)|^2
\int dm' \, |\rho(m')|^2
e^{-i(m-m')t/\gamma}
.
\label{e43}
\end{align}
Finally,
taking into account the normalization in Eq.~(\ref{e006}),
from Eq.~(\ref{e039}) we obtain
\begin{equation}
P_{\vet{v}}(t)
=
\int dm \, |\rho(m)|^2
\int dm' \, |\rho(m')|^2
e^{-i(m-m')t/\gamma}
.
\label{e44}
\end{equation}
This expression for the survival probability clearly satisfies the
relativistic relation (\ref{e010}),
as we wanted to prove.

\section{Conclusions}

In the derivation of the survival probability (\ref{e44})
we assumed the conditions (\ref{e031}) and (\ref{e421}):
\begin{equation}
\Gamma \ll \sigma_{p} \ll M
\quad
\Longleftrightarrow
\quad
M^{-1} \ll \sigma_{x} \ll \tau
.
\label{e45}
\end{equation}
Let us emphasize that the conditions
$\Gamma \ll M$
and
$\Gamma \ll \sigma_{p}$
are necessary for the measurability of the decay law of an unstable system
and
the condition
$\sigma_{p} \ll M$
is necessary for the existence of an inertial reference frame in which the system is at rest.

The conditions (\ref{e45}) are verified in all experiments
which measure the lifetime of unstable systems\footnote{
The conditions (\ref{e45}) cannot be satisfied in the case of meson and baryon resonances
in high-energy experiments,
for which $\Gamma \sim M$.
However,
in this case the decay law as a function of time is not measurable.
}.
For example,
let us consider the decay of a charged pion,
which has a small mass
$M(\pi^{\pm}) \sim 10^2 \, \text{MeV}$
and a large decay rate
$\Gamma(\pi^{\pm}) \sim 10^{-8} \, \text{eV}$.
In this case
$M^{-1}(\pi^{\pm}) \sim 10^{-13} \, \text{cm}$
and
$\tau(\pi^{\pm}) \sim 10^{3} \, \text{cm}$.
If the pion decays at rest in matter,
its spatial uncertainty is of the order of the interatomic distance,
$\sigma_{x} \sim 10^{-8} \, \text{cm}$,
and
the conditions (\ref{e45}) are very well satisfied.
Other nuclear and atomic systems have larger mass and longer lifetimes,
which satisfy the conditions (\ref{e45}) even better
if the system is localized with a realistic uncertainty that can go from
the interatomic distance to a few centimeters.

In conclusion,
we have shown that the conditions (\ref{e45}),
which are satisfied in all experiments
which measure the lifetime of unstable systems,
lead to the standard relativistic relation
(\ref{e010})
between the survival probability
of an unstable system in motion
and that of the system at rest,
that follows from elementary considerations
based on relativistic time dilation.
This result
has been obtained with a relativistic
approach which takes into account not only the time evolution,
as done usually,
but also the space evolution of an unstable system.
Therefore, we confute the claims of non-standard quantum-mechanical
effects presented in
Refs.~\cite{Shirokov:2004gi,Stefanovich:2005ai,quant-ph/0508087,physics/0603043,Urbanowski:2014gza},
which are in clear contradiction with Special Relativity.

\acknowledgments

We would like to thank K. Urbanowski for stimulating discussions.
S. A. Alavi would like to thank the
Department of Physics of the University of Torino for hospitality during the realization of this work.
S. A. Alavi would also like to thank Mr. Ali Dowlatabadi for his kind assistance.
S. A. Alavi is supported by the Ministry of  Science, Research and Tecknology of Iran and also  by the Hakim Sabzevari University.
The work of C. Giunti is supported by the research grant {\sl Theoretical Astroparticle Physics} number 2012CPPYP7 under the program PRIN 2012 funded by the Ministero dell'Istruzione, Universit\`a e della Ricerca (MIUR).

\end{document}